# Dynamic Index NAT as a Mobility Solution in OMNeT++

Atheer Al-Rubaye, Jochen Seitz
Communication Networks Group
Technische Universität Ilmenau, Germany
{atheer.al-rubaye, jochen.seitz}@tu-ilmenau.de

*Abstract*—**Mobility in wireless networks causes a major issue from the IP-addressing perspective. When a Mobile Node (MN) moves to another subnet, it will probably get assigned a new IP address. This causes a routing problem since the MN will not be reachable with its previous IP address known to the other communication party. Real time applications might suffer from connection drops, which is recognized as inconvenience in the currently used service, unless some solution is provided. An approach to maintain session continuity while traversing heterogeneous networks of different subnet addresses is proposed. Here, a cross-layer module is implemented in OMNeT++ with NAT functionality to provide a seamless handover. A proof of concept is also shown with analogy to the Mobile IPv6 protocol provided in INET.**

## I. Introduction

Wireless communication networks can be categorized as heterogeneous based on different aspects, like having an infrastructure, type of radio access technology, and subnet address. While nowadays communication devices are equipped with multiple interfaces and due to the evolving applications and usage scenarios that require IP connectivity anywhere anytime, switching the connection to another access point (handover) and getting assigned a new IP address is a common communication scenario.

In this paper, we try to handle this issue from the network layer point of view and observe its impact on running applications and the provided Quality of Service (QoS). Nevertheless, we also present our design for vertical handover (VHO) management module, which is however not yet provided with performance analysis. The module manages VHO from above the link layer and between independent wireless interfaces.

For a running session, changing the IP address introduces a routing problem where the MN is not available any more for its communication party, unless a robust mobility solution is deployed. Our proposed approach makes use of the fact that due to the limited address space of IPv4 and the increased number of IP-connected users, Network Address Translation (NAT) became a de facto standard in almost all communication networks. This work implements our concept previously presented in a use case in [1]. Also it contributes to the INET framework of OMNeT++ [2] by implementing NAT operation in network layer with an update mechanism achieved through a cross layer module as will be described in more details later.

Performance results are provided in comparison to the known mobility solution (MIPv6) provided in INET. The program code introduced here is a part of a PhD research work and can be available for sharing after the defense of the PhD, to be presented as a VHO protocol contributing to the INET framework.

## II. Related work

Next generation network is predicted and being realized to be IP-based, thus service continuity while handover through IP address resolution is a sophisticated research field in wireless communications. The well known mobility protocols Mobile IPv4 (MIPv4) and Mobile IPv6 (MIPv6) [3] [4] are standardized as mobility solutions for the two versions of IP protocols. The idea is to achieve a transparent routing of IP datagrams independently from the mobile host location. This however, introduces delay due to the use of non optimal routing through the home agent, and requires the correspondent node to support the protocol in case route optimization is to be carried out. Several extensions like Fast MIPv6 [5], Hierarchical MIPv6 [6] and Proxy MIPv6 [7] have been proposed that extend the basic MIPv6 to offset its disadvantages. However, IPv6 is not dominant yet in all of our networks neither expected to come fast, and we have to still live with IPv4 for pretty much of time. We assume that the MNs are the initiators of their sessions with the requested application server to start a video stream while the MNs are mobile. However, to have them also reachable for other session initiators from outside, the same update scheme can be deployed to inform the corresponding DNS servers in the area. In this case, private IP addressing is not feasible any more and then, public IP addressing is a necessity. To solve this, the approach of [8] uses a global naming scheme above the IP layer. Although it can be useful for the case of vehicular mobile adhoc communications, it might however harm delay sensitive applications. In relevance to our VHO framework, the one presented in [9] focuses on modeling two types of wireless interfaces and compares two decision algorithms, but a node is not actually able to perform a VHO principally, since the decider node selects a path to send the traffic through, which is a path selection in concept rather than a handover to a new sender/receiver interface of an MN. Our work suggests a quick solution for today's networks of IPv4 with minimum changes to be applied to existing protocols and nodes keeping the problem and the solution as close to the user as possible. To have adequate performance analysis and comparison, we prove our concept in analogy to MIPv6 that is already provided in INET. At least we do not expect it to provide worse performance than MIPv4, so our comparison should still be valid when we argue upon solutions of different IP versions here. The network setup used is similar to the one presented in [10], which is implemented in INET as well to show MIPv6 functionality. Our paper is in fact a part of bigger research work in progress that proposes a vertical handover cross layer-based engine within the INET framework. So far, in our mobility solution we suggest basically a NAT server capable of dynamically updating its entries based on updates generated by the MN at handover. Namely, we initiate another mobility solution in OMNeT++/INET, try to show the pros and cons of such an approach and explore its feasibility.

### III. ADDRESSING CONCEPT

We call our approach Dynamic Index NAT (DINAT) to differentiate it from the well known dynamic NAT. In traditional setups of networks deploying local IP addresses, a gateway with NAT functionality creates a new entry containing global/local IP addresses and ports for any new hosted MN. Any session with a node outside the local network will use the global IP address/es of the gateway. If a handover takes place, a session drop is a major consequence and the MN will then need to restart any session that was running because packets will still be destined to the old IP address of the MN. In our scheme, we assume as a primary step that the different networks are coupled through a single gateway we call GRouter that provides NAT, as depicted in figure 1. This is however a kind of restriction, but we want to proof the feasibility in this stage, and then relax it later in the form of a proxy server available in the cloud as an anchor point for MNs communications.

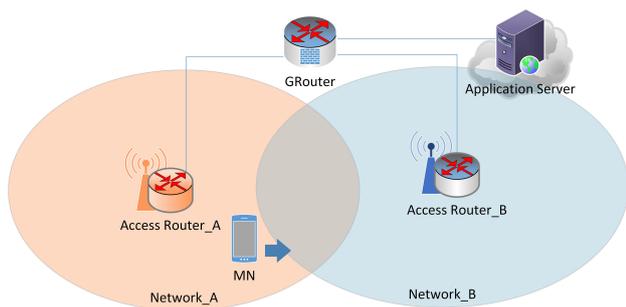

Fig. 1. Network topology

As long as an MN is within a certain network, it is provided with normal NAT functionality by the GRouter. When the IP address of a MN is changed after a handover, a chain of updates between the MN and the gateway of the network takes place. The GRouter receives an update message from the MN including its old and new IP settings to update its NAT entries, so all incoming packets of a running session will be translated to the recently assigned IP address and hence, packets will be re-routed to the new network. Port Address Translation (PAT) is also implemented to support more clients in case of IP overloading in NAT. Figure 2 demonstrates the DINAT mechanism.

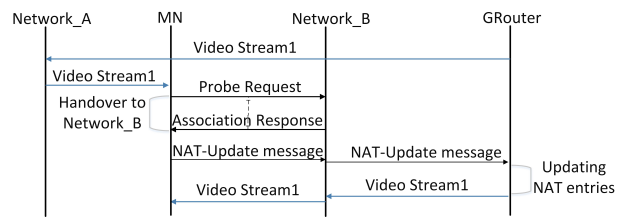

Fig. 2. chart of exchanged messages

### IV. SIMULATED MODULES

The suggested approach targets to present a VHO solution within the INET frame work. To realize the design in OMNeT++, functions are added at the Network and Access layers, where both are connected to a cross layer module. We categorize and describe each briefly as follow:

#### A. Access layer related and cross layer modules

A cross layer module we call the IPCoManager is implemented in the WirelessHost/MN. It manages the wireless interfaces in VHO operations through a module called VHO-Controller. This controls the modules in the wireless interfaces that govern association process with their corresponding networks. Our approach does not care about the details of the differences between the wireless technologies for now but rather, governs their rules in connectivity with respect to the upper layers. A primitive request message will always be sent by the controlling module of an interface (by the Agent module of the WLAN for ex.) to the VHOController to acquire permission to connect. The decision depends on the state of the controller and the result of the decision algorithm inside. Figure 3 illustrates the controller phases of operation from a link layer prospect.

For a stable handover, the controller implements a set of timers that can be setup in the simulation configuration file. For example, when switching ON the user device, the module waits for a specific time to make sure that all surrounding beacons are received and hence selects the best (if no preferred is specified), rather than immediately associating with the first discovered network. This timer is configurable according to the beaconing intervals. A new request is denied if a VHO has just been carried out or a permission is currently waiting for confirmation of association. A time of dual connectivity mode is also possible before releasing the old connection. On a grant of permission to connect and eventually the receipt of Association Confirmation from the concerned interface, the IPCoManager generates NAT update messages that contain old and new IP addresses of the MN before and after VHO respectively. The generated message is then sent to the IPv4 entity for further process. A local version of the message is also generated when switching ON the device and first

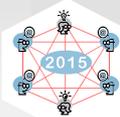

selecting a hosting network to set the active interface and gateway at the network layer.

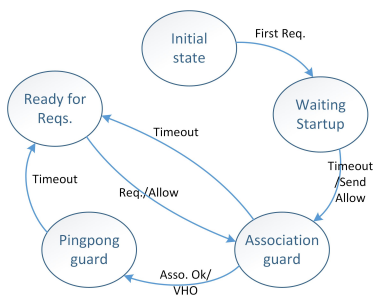

Fig. 3. VHOController state diagram

### B. IPv4 related modules

At the host (MN) side, the IPv4 module receives a NAT update message from the IPCoManager in case of handover. It updates its routing table according to the IP address information carried. The NAT update message is then forwarded to its specified destination, which is the GRouter in our case. A local version of the message is received every time the user device is switched ON to set the active interface and gateway in the routing table in accordance to the selected network, and then it is discarded. To add NAT functionality to our GRouter, the IPv4 module provided in INET is deployed. However, to maintain the flexibility of INET, the new version of the IPv4 is deployed through a module interface, since it is required only for the GRouter to provide NAT. A simple module we call NAT-Table is implemented such that, whenever a data packet is received in IPv4, NAT-Table is consulted and the respective IP addresses and Port numbers (for NAT and PAT) are changed accordingly. A new message type and a handle function are defined in IPv4 in relevance to NAT functionality to make it capable of amending its entries dynamically. Whenever the message called NATUpdate is received at IPv4, the information carried by the message (old and new IP/port address) are used to update the specified entry in NAT-Table. Figure 4 shows the added modules.

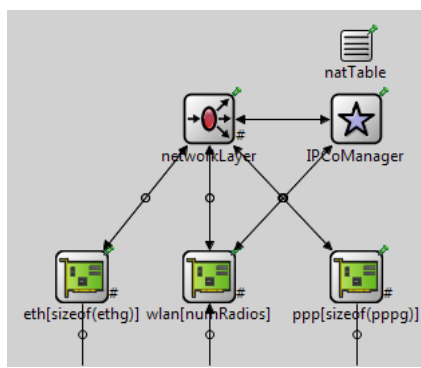

Fig. 4. DINAT modules

## V. SIMULATION SETUP AND SCENARIO

The topology of figure 1 demonstrates two networks of different subnet addresses with an MN moving from one network to the other, which is most likely the case for networks of different radio technologies as well. In relevance to the modules description in the previous section, the simulation scenario considers a single wireless interface in the MN. Since this work emphasizes only the addressing problem, radio technology and vertical handover related tasks at the link layer are out of the scope here. The mobile node moves in a linear path from a predefined point inside the coverage area of Network_A, in which it is associated first, towards a specific point within the coverage area of Network_B, where it handovers its connection to. In the meanwhile, it requests the Application Server to run a video stream as a real time application. For proof of concept purpose, the client/MN is enabled to re-request streaming after specific time of stream receipt absence (we call App. Time out). In general, all the IP addresses are assigned statically in the simulation. In different simulation runs, Mobility speed values of 1,2,10 and 20 mps were tested with sending intervals of 5,10,100 and 500 ms at the Application Server using the same packet size. Throughout each run, the MN moved at a constant speed and on a linear path using the mobility module LinearMobility of INET. The simulation time differs for each run according to the mobility speed, however, if equal time was used for all the runs, we would have the MN moving back and forth between the simulation boundaries when high mobility speed is set and hence, more handovers would be experienced, which in turn will produce inconsistent results. Position of all nodes and mobility speed are predefined, so no simulation repetitions were performed for a single set of parameters, but rather for different sets. MIPv6 main parameters like router advertisement intervals are left as in the INET defaults.

To investigate the feasibility of our approach, we compare the impact of handover on two more scenarios in addition to ours that we call DINAT _Case. The second scenario we call Default_Case, which uses absolutely no mobility solution other than a reaction for the stream interruption at the application layer by re-requesting the stream. MIPv6_Case is our third scenario, which applies MIPv6 provided by INET. The three scenarios use the topology shown in figure 1 with the same set of simulation parameters.

## VI. RESULTS AND ANALYSIS

Since the worse consequence for a handover is the unreachability of the MN, we measure the performance here in terms of packet loss rate in each scenario. This loss rate was measured in multiple simulation runs as an average versus MN mobility speed, where for each velocity value a different set of sending interval values for the Application Server were used. Figure 5 shows the performance of the three scenarios, each ran with a single handover event.

Since the default case has no mobility solution and relies only on the application layer to react after a time of no-receipt of packets is reached, this scenario shows the highest packet loss as expected and observed through multiple simulation runs. As identification aspect, this is considered as a new stream or session setup at the server. Showing results for a scenario with no mobility solution might not be common but, at least it expresses the amount of improvements the different tested solutions brought. MIPv6 shows a significant improvement over the last. The MN is still defined with its home IP address at the server (and with its new one if we include route optimization procedure) and session continuity is achieved. Our DINAT shows a further improvement, where it is not depending on router advertisements as MIPv6. The handover and re-routing delay depends on the handover delay time, which is link layer related, and on the time starting from trigger initiation at the link layer of the MN till the NAT table in the GRouter is updated. Mobility speed has a huge impact on the performance overall as illustrated, too.

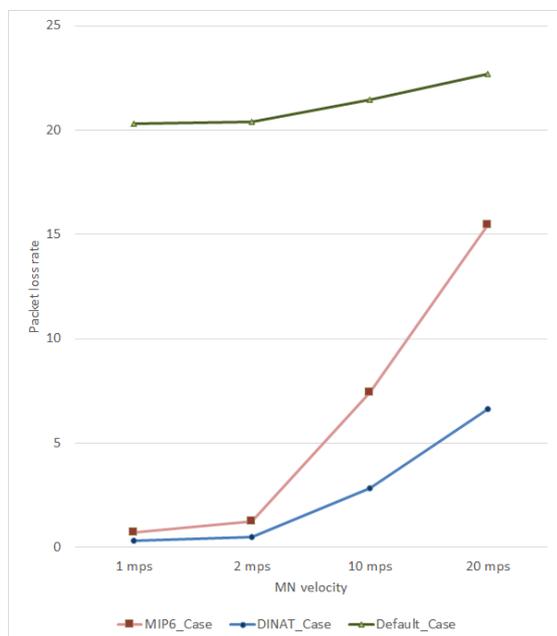

Fig. 5. Packet Loss Rates

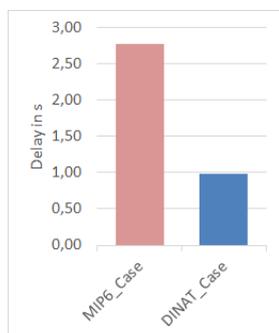

Fig. 6. Delay experienced at the Application layer

If we like to consider the delay sensed by the running application, which we define as the time between the last and the first packets were received before and after the handover respectively, then figure 6 can demonstrate the delay for two of the tested scenarios. There was no point to show that of the Default_Case since it equals to the client App. Time out. However, we can observe less delay experienced by the application in the DINAT_Case.

## VII. CONCLUSION AND FUTURE WORK

We presented a solution to deal with addressing problem as a consequence to handover between different subnets. The proposed approach showed a better performance as it was tested with a video stream traffic. The benefits were shown in terms of packet loss rate and delay sensed at the application layer. Randomness was not considered in the simulation runs, where we focus only on the behavior during the handover and its impact on the running traffic. The well known MIPv6, which is already implemented in INET, was chosen as an opponent approach. Despite it deals with a different category of IP addresses, we expect that MIPv4 would not perform better than MIPv6. Modified versions of MIPv6 are not available for MIPv4, so our comparison will still hold if we want to consider a solution for todays IPv4-based dominant communication networks. In comparison to MIPv6, our approach shows a better performance in term of packet loss during handover, deploys the standard version of IP without the need to add any mobility extension to IP in all the nodes between the MN and the corresponding node, and requires no real IP assignment to the MN. TCP traffic was however not in the scope, since such a type of traffic is less sensitive to handover and packet loss. No modification is needed at application server side, which was intended to be masked from all handover signaling and subsequent events. The NAT update message that is deployed to inform the GRouter about the IP address change however, is not transmitted yet in a reliable manner, so any loss or delay will cause a big negative impact on the overall performance. The network topology tested might not be common in real networks for reasons related to management, quality of service and security, but this work aims to upgrade the DINAT functionality to a proxy server setup globally as an anchor point for the traffic, without the need to couple the network like we did through the GRouter. Users who participate to a certain service will forward their packets to the proxy. Added delay, which may become large then, is an open question on the feasibility of such a setup. The scenarios tested here apply no processing load to the GRouter since only one MN was running. Future work should take into account observing the performance within networks loaded with users. In addition to the mentioned future tasks, the proposed approach is to be expanded to present a package solution for vertical handover. In this case, signaling at the link layer regarding multiple independent wireless interfaces, the decision algorithm concerning handover initiation and network selection, and an address resolution mechanism (like DINAT) represent challenges to be considered.